\documentclass[fleqn,usenatbib]{mnras}

\usepackage{newtxtext,newtxmath}

\usepackage[T1]{fontenc}
\usepackage{ae,aecompl}
\usepackage[ampersand]{easylist}

\usepackage{graphicx}	
\usepackage{amsmath}	
\usepackage{amssymb}	
\usepackage{subfigure}  
\usepackage{xcolor}
\usepackage[flushleft]{threeparttable}
\usepackage{verbatim}

\newcommand{\kepler}{{\it Kepler}}

\newcommand{\kms}{km\,s$^{-1}$}

\newcommand{\mjup}{\mbox{M$_{J}$}}

\newcommand{\msun}{\mbox{M$_{\odot}$}}

\newcommand{\mearth}{M$_{\oplus}$}

\newcommand{\feh}{[Fe/H]}

\newcommand{\hJ}{hot Jupiter}
\newcommand{\hJs}{hot Jupiters}

\newcommand{\bes}{Besan\c{c}on}
\newcommand{\pmc}{planet-metallicity correlation}
\newcommand{\sweet}{SWEET-Cat}

\title[Hot Jupiter PMC]{Investigating the Planet-Metallicity Correlation for Hot Jupiters}

\author[A. Osborn]{
Ares Osborn$^{1,2}$\thanks{E-mail: e.osborn@warwick.ac.uk}
and Daniel Bayliss$^{1,2}$
\\
$^{1}$Department of Physics, University of Warwick, Gibbet Hill Road, Coventry CV4 7AL, UK\\
$^{2}$Centre for Exoplanets and Habitability, University of Warwick, Gibbet Hill Road, Coventry CV4 7AL, UK\\
}

\date{Accepted XXX. Received YYY; in original form ZZZ}

\pubyear{2019}

\begin{document}
\label{firstpage}
\pagerange{\pageref{firstpage}--\pageref{lastpage}}
\maketitle

\begin{abstract}
We investigate the giant planet-metallicity correlation for a homogeneous, unbiased set of 217 hot Jupiters taken from nearly 15 years of wide-field ground-based surveys.  We compare the host star metallicity to that of field stars using the \bes\ Galaxy model, allowing for a metallicity measurement offset between the two sets. We find that hot Jupiters preferentially orbit metal rich stars.  However, we find the correlation consistent, though marginally weaker, for hot Jupiters ($\beta=0.71^{+0.56}_{-0.34}$) than it is for other longer period gas giant planets from radial velocity surveys.  This suggests that the population of hot Jupiters probably formed in a similar process to other gas giant planets, and differ only in their migration histories. 
\end{abstract}




\section{Introduction}
\label{sec:intro}
It only took the first few discoveries of giant exoplanets to notice that the host stars have a higher metallicity content compared with field stars hosting no planets \citep{Gonzalez1997,Santos2000,Santos2001} - \citet{Gonzalez1997} proposed this link after just four giant planets were detected. This result has evolved into the now well-known giant planet-metallicity correlation; that is, the higher the metallicity of a star, the more likely it is to host a giant planet \citep{Santos2004,Fischer2005,Johnson2010,Maldonado2012,Mortier2013,Schlaufman2014}.  This result was recently reviewed by \citet{Adibekyan2019}, who reanalysed the giant planet-metallicity correlation using the homogeneous stellar parameters listed in the \sweet\ catalogue \citep{Santos2013}. They contrasted their sample of FGK dwarf star hosts (with planets discovered by the RV and transit methods) with a comparison sample of FGK stars hosting no planets from the HARPS GTO program \citep[see][]{Adibekyan2012}, which has stellar parameters derived using the same method as those in \sweet\ (thus making them directly comparable). They show a very obvious difference in the distribution of metallicity of stars without planets compared to stars hosting giant planets (see fig.~5 in \citealt{Adibekyan2019}), confirmed with a two-sample Kolmogorov-Smirnov (KS) test and reaffirming the existence of the giant planet-metallicity correlation. 

When the giant \pmc\ was established, there were two main theories proposed as to why it occurs. The first, pollution or self-enrichment (suggested as the mechanism behind the correlation in \citealt{Gonzalez1997}), suggests that the outer convective envelope of the star is polluted by an infall of material onto it, perhaps due to the inward migration of a gas giant planet. Primordial origin is the second: in this, the metallicity of the star is representative of the metallicity of the primordial cloud from which the star formed. This would imply that, in a high metallicity protoplanetary disc, giant planets form more easily. \citet{Santos2001} concludes that a simple pollution model cannot be the key process that leads to the metallicity offset of stars with planets, and primordial origin is further corroborated by results from \citet{Santos2003,Santos2004,Valenti2005,Valenti2008,Johnson2010}; and \citet{Maldonado2012}. It is also supported by the core accretion planet formation theory \citep[e.g][]{Ida2004a}, one of the leading theories of planet formation.

Core accretion \citep[e.g.][]{Pollack1996} is a bottom-up process, wherein the formation of giant planets begins with a rocky/icy core (10-15 \mearth); gas is then accreted onto the core in a runaway process until it has either cleared its orbit or the gas has been removed from the disk. If the initial disk has a higher metallicity content (i.e. more grains), it is expected that the large metal cores that go on to efficiently accrete gas would be more easily built, before the gas in the disk is lost. The \pmc\ is thus an important piece of observational evidence in support of this scenario. Core accretion timescales were thought to be longer than the lifetime of the disk, but have since been found realistic when including disk evolution and migration \citep[e.g.][]{Rice2003a,Alibert2004}. \citet{Adibekyan2019} suggests that a \textit{combination} of longer disk lifetime \citep[e.g.][]{Ercolano2010} and the presence of more material to form cores \citep[e.g][]{Mordasini2012} that results from a  higher metallicity protoplanetary disk can both influence the formation and migration of giant planets. \citet{Ida2004b} and \citet{Benz2006} suggest that in a high metallicity environment, giant planets could form more efficiently (allowing more time for migration) and/or closer in to the star, potentially inside the snow line. 

This study looks at the giant planet-metallicity correlation for a subset of giant planets on short period orbits: namely the "\hJ" planets. The first discovery of an exoplanet around a main sequence star was the \hJ\ 51 Peg b \citep{Mayor1995}, and since then they have been found in their hundreds. Despite being relatively easy to find with transit and radial velocity surveys, due to their large radii and short orbital periods, \hJs\ are comparatively rare, with occurrence rates around FGK type stars of 0.4\% \citep{Cumming2008,Howard2012,Zhou2019}. 

We compare the giant planet-metallicity correlation of hot Jupiters to giant planets with longer periods, such as those found by the radial velocity surveys of \citet{Valenti2005} and \citet{Schlaufman2014}.
Few papers have looked at the \pmc\ for \hJs\ in particular, but some (sometimes contradictory) trends have been observed. 

\citet{Sozzetti2004} shows a lack of planets on very short period orbits ($\leq 5$ days) around stars with a metallicity less than solar, but due to potential biases and small-number statistics, cannot draw a clear conclusion. Some years later, with an increase in the number of hot Jupiter discoveries, \citet{Maldonado2012} found that at lower metallicities, hot giant planets are less frequent than their cool giant counterparts. \citet{Adibekyan2013} shows that planets (from 0.03\mjup\ to 4\mjup) around metal-poor stars have longer periods. But \citet{Narang2018} observes that, while the average metallicity of the host star increases for orbital periods of less than 10 days when planets of up to 50 \mearth are present, there is no disparity between the average metallicity of stars hosting short ($\leq 10$ days) and long ($>10$ days) period giant ($>50$ \mearth) planets. 

Returning to \citet{Adibekyan2019}, it is now worth noting that a KS test shows the hosts of their separate radial velocity and transiting planet samples have indistinguishable metallicity distributions, despite the planets having significantly different orbital period regimes. Their transiting sample has an average orbital period of 11 days, whereas the average for their RV sample is 1202 days. Unfortunately, the average of the transit sample in \citet{Adibekyan2019} is a little over the 10 day threshold that defines a short period giant planet in \citet{Narang2018}, therefore making the two incomparable.

In this paper we begin in Section~\ref{sec:sample} by defining a sample of homogeneous transiting hot Jupiters that have been detected from wide-field, non-targeted, transit surveys. In Section~\ref{sec:analysis} we compare this sample to a distribution of field stars drawn from the \bes\ Galaxy model.  In Section~\ref{sec:results} we set out our findings in the context of the planet-metallicity correlation for hot Jupiters, and in Section~\ref{sec:discussion} we discuss these results in the context of previous surveys and planet formation theory.  Finally, we set out our conclusions in Section~\ref{sec:conclusion}.

\section{The sample of hot Jupiters} \label{sec:sample}
 As their name suggests, hot Jupiters are exoplanets with masses and radii similar to Jupiter, but in very short (hot) orbits around their host stars.  51 Peg b \citep{Mayor1995} is an archetypal hot Jupiter.  The precise definition of a \hJ\ varies a little in the literature; in this study we define it as an exoplanet with mass between 0.1 and 13 \mjup\ (the upper limit being the approximate mass at which deuterium burning becomes possible; the lower limit ensures that the planets within the sample are gas giants and not terrestrial), and a period of up to 10 days (inclusive). 

In order to probe whether the planet-metallicity correlation is different for \hJs\ in comparison to longer period gas giants, we have compiled a sample of confirmed transiting \hJs\ taken from the NASA Exoplanet Archive\footnote{https://exoplanetarchive.ipac.caltech.edu/} \citep{Akeson2013}.  To ensure we have a sample free from any biases, we only select exoplanets which have been discovered from non-targeted surveys - i.e. from wide-field surveys where all stars within the field-of-view are searched.  This naturally excludes any radial velocity discoveries, and also surveys such as \kepler\ \citep{Borucki2010} and \textit{K2} \citep{Howell2014}, where only pre-selected stars were monitored.  However, it does include the vast majority of hot Jupiter discoveries, and these discoveries have predominantly originated from the wide-field ground-based surveys:  WASP \citep{Pollacco2006}; HATNet \citep{Bakos2004}; HATSouth \citep{Bakos2013}; KELT \citep{Pepper2007}; XO \citep{Crouzet2018}; and TrES \citep{Alonso2004}.  This unbiased sample is required in order to compare metallicity of hot Jupiter hosts to that of field stars drawn from a synthetic galaxy population such as \bes.  We also removed any \hJ\ in a system with more than one star, either confirmed on the NASA Exoplanet Archive or suggested in its discovery paper, as previous literature has shown that stellar binaries \citep{Eggenberger2004,Eggenberger2011} and stellar multiplicity \citep{Wang2014a,Wang2014b} have an effect on planet formation \citep[as summarised in][]{Wang2015}, and we did not wish to unintentionally bias the sample.  Finally, we also exclude a small number of host stars with visual magnitudes of 9 or brighter, as for such systems we could not generate a large enough field stars distribution from the \bes\ Galaxy model.

\begin{figure*}
    \centering
    \subfigure{\includegraphics[width=0.45\textwidth]{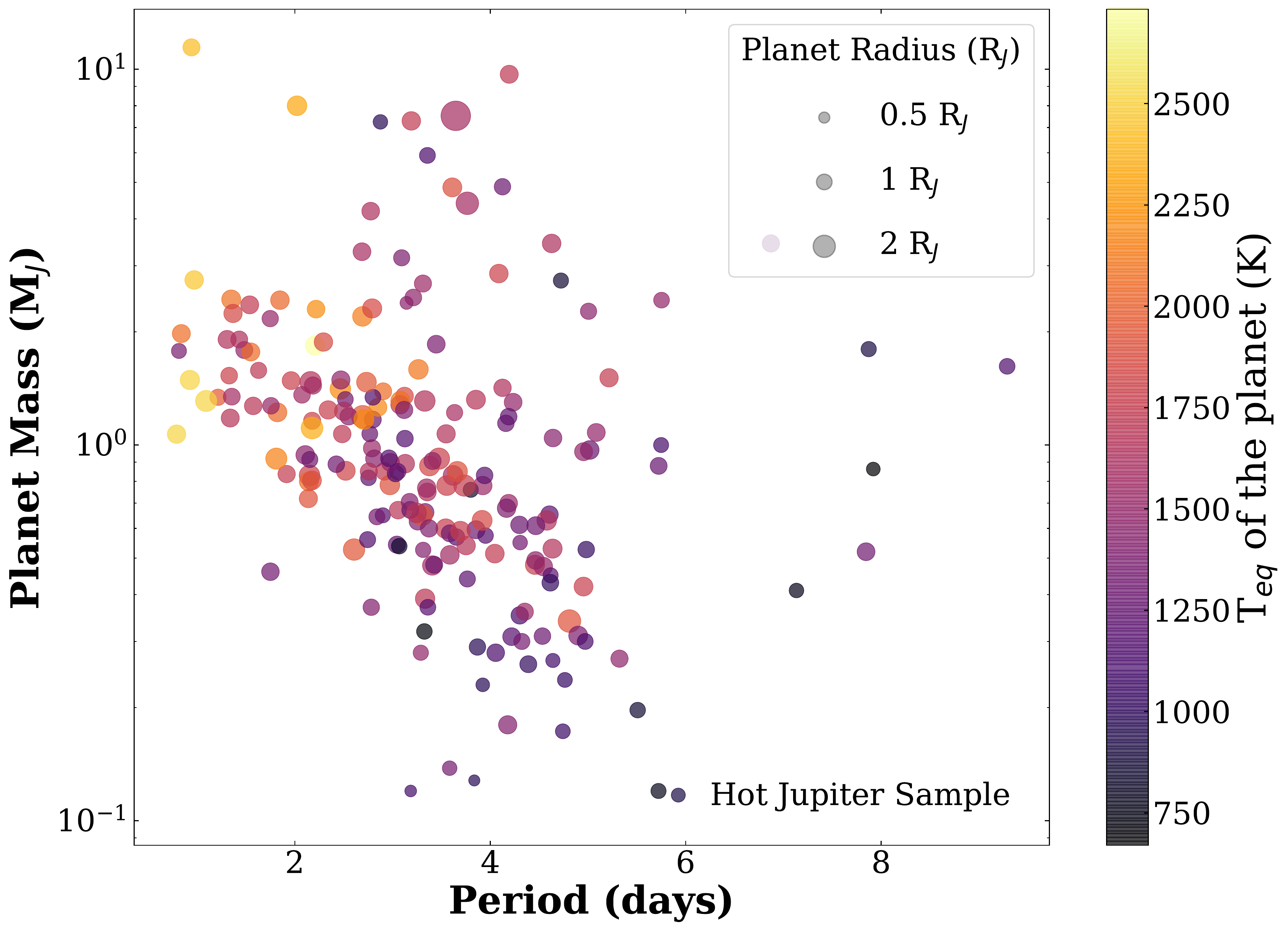}} 
    \subfigure{\includegraphics[width=0.45\textwidth]{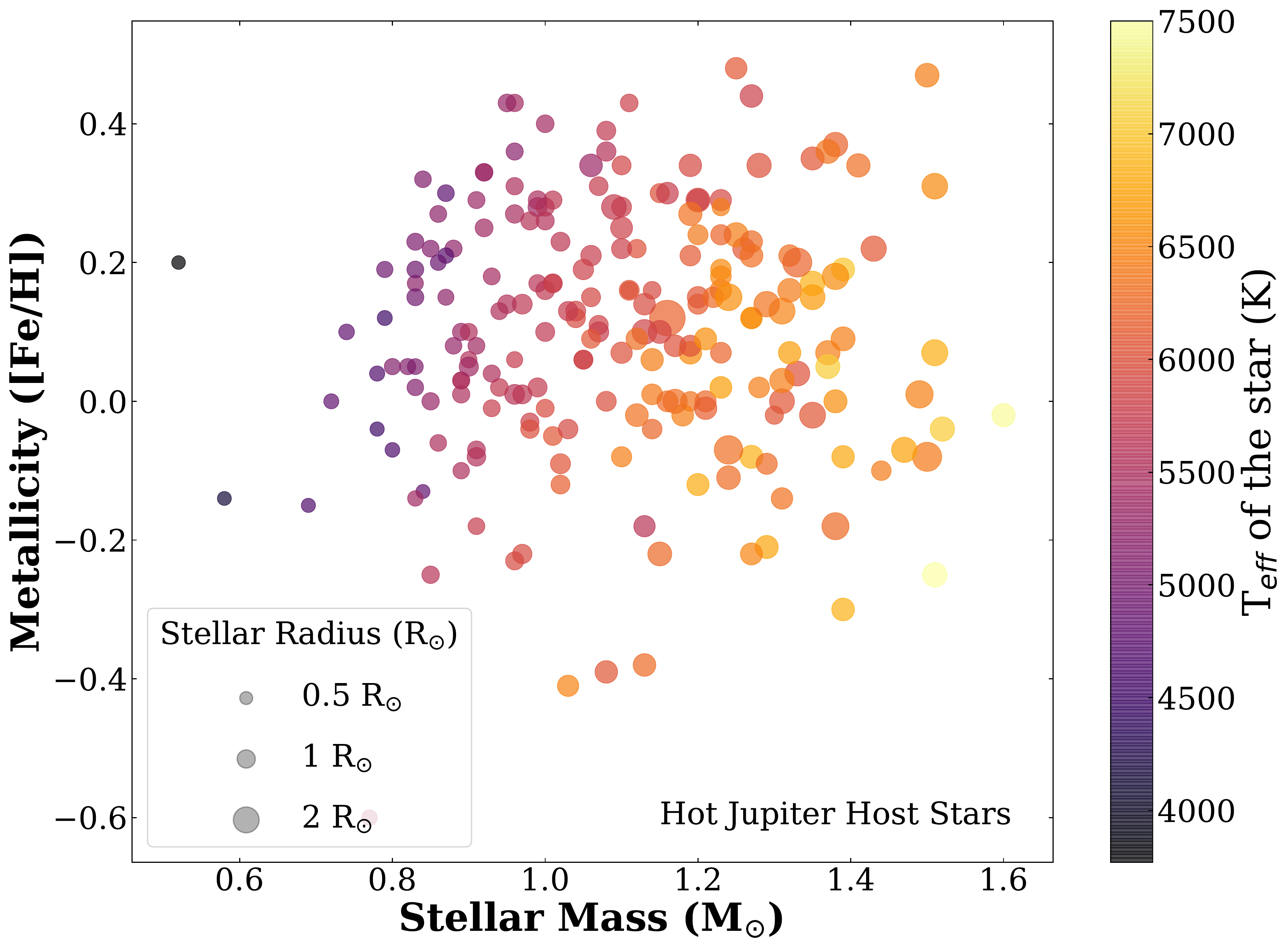}}
    \caption{Properties of the 217 transiting \hJs\ in our sample. Left: the \hJ\ sample, displaying the mass and period selection criteria of 0.1 to 13 \mjup\ and up to 10 days respectively. Planet radius scales with the size of the marker (a larger marker indicates a larger radius), and equilibrium temperature scales with the marker colour (where yellow is hotter, and \textbf{purple} is cooler). Right: the metallicities and masses of the host stars of the hot Jupiter sample, with properties taken from \sweet. Similarly, star radius scales with marker size, and effective temperature with marker colour.}
    \label{fig:planetandstarprops}
\end{figure*}

Our final sample consisted of 217 \hJs, each with a corresponding host star.  We present the properties of these hot Jupiters and their host stars in Figure~\ref{fig:planetandstarprops} (left and right respectively), with parameters taken from the \sweet\ catalogue \citep{Santos2013}.

Metallicity is commonly expressed in terms of \feh, which is the logarithm of the ratio of a star's iron abundance compared with the Sun.  The element iron is used due to strong, numerous, and easily measurable iron lines in the optical spectra of solar-type stars.  We adopt \feh\ for this study, as it allows us to easily compile a homogeneous set of metalliticities for our host stars, and it means we can easily compare our work to previous studies. Metallicities for the hot Jupiter host stars were taken from the \sweet\ catalogue of stellar parameters \citep{Santos2013}, as this catalogue is the largest collection of host star and planet parameters that have been derived in a homogeneous way. \sweet\ utilises high resolution and high signal-to-noise spectra in deriving stellar properties, with a uniform methodology based upon the principles of iron ionisation and excitation equilibrium. Different groups use different analysis methods to derive their metallicities, which can introduce significant offsets - for example, \citet{Torres2012} shows that the difference in average metallicity calculated independently by the WASP and HATNet groups for a comparable samples of stars is about 0.17 dex. By using metallicity values solely from \sweet, we circumvent this issue. 


\section{Analysis}\label{sec:analysis}

\subsection{\bes\ simulation of field star populations}
In order to determine if there is a correlation between the occurrence of \hJs\ and host star metallicity, we need to determine the distribution of stellar metallicities from which the \hJ\ host star was drawn.  However, typically for a transit survey field, high resolution spectroscopy capable of determining metallicity is only undertaken on transiting planet candidates.  This means that the overwhelming majority of stars in the transit survey do not have measured metallicities.  It is therefore necessary to calculate the distribution of stellar metallicities from a simulated ``field'' star population from which our sample of hot Jupiter host stars was drawn.  To make this simulation, we use the 2003 \bes\ Galaxy Model \citep{Robin2003}, which provides the metallicity for individual simulated stars (in \feh) in a given parameter range \citep{Robin1996}.

We performed the default \bes\ Catalogue simulation without kinematics, using the Johnson-Cousins Photometric system. We created a population of simulated stars for each individual \hJ\ host star, with a range of galactic latitude and longitude within 10 deg$^2$ of the \hJ\ host star.  The simulated population was restricted to stars with a visual magnitude with $\delta V_{\rm mag} = 1$ of the \hJ\ host star.  We further restricted the population to dwarf stars, which removed distant giants for which transit surveys were not sensitive to finding hot Jupiters around.  Finally, we restricted the population to the mass range of the host star sample, from 0.52 to 1.6 $\msun$, in order to remove very high mass stars, which again transit surveys were not sensitive to finding hot Jupiters around.  All other \bes\ Galaxy Model parameters were kept to the default values, as these have been shown to best simulate stellar populations in our local Galaxy when compared with large spectroscopic surveys \citep{nandakuma2017}.
For each of our 217 \hJ\ host stars, we selected 50 stars from the corresponding simulated stellar populations which were closest in visual magnitude value to the \hJ\ host star.  Thus our final set of simulated stellar populations comprised of 10850 stars in total. 

\begin{figure}
	\includegraphics[width=\columnwidth]{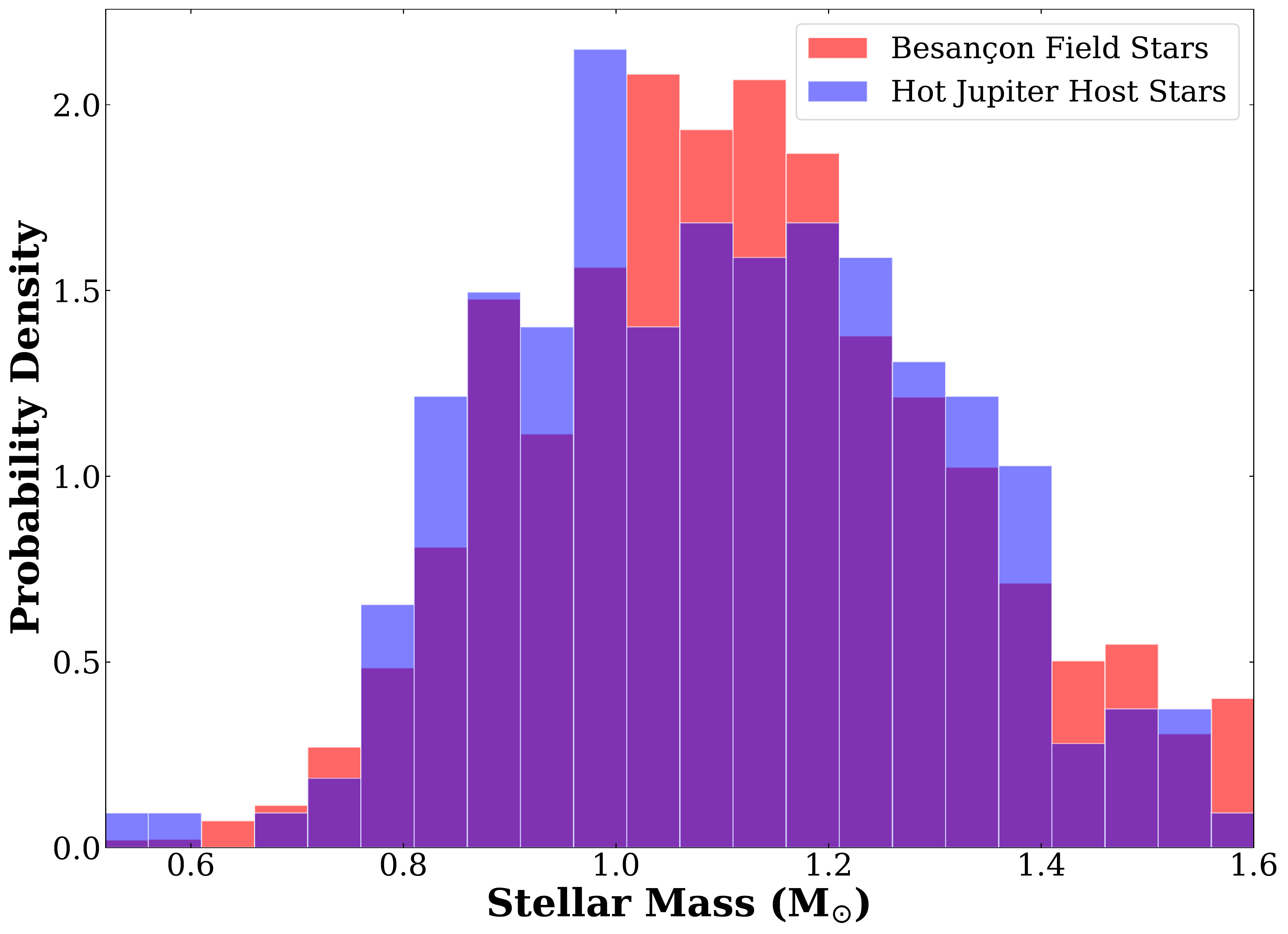}
    \caption{A comparison of the mass distribution of the host stars belonging to the hot Jupiter sample (blue), and the field stars simulated by the \bes\ galaxy model (red). The field star sample has been cut to only include masses within the range of the host stars (0.52 to 1.6 $\msun$), and to only include 50 field stars of the closest visual magnitude to each host star.  The masses from the model stars are in good agreement with the hot Jupiter host star masses.}
    \label{fig:massoffieldandhost}
\end{figure}

In order to examine whether our simulated stellar populations accurately represent our sample of \hJ\ hot stars, we make a comparison of the stellar mass distributions of each - see Figure ~\ref{fig:massoffieldandhost}.  Performing a KS test on the two distributions returns a statistic of 0.071 and a p-value of 0.23, indicating that the masses of the \hJ\ host stars are likely to be drawn from the same population as the simulated \bes\ stellar population.  This gives us confidence that the simulated \bes\ population does represent the stellar population from which the \hJ\ host stars are drawn.

\begin{figure*}
	\centering
	\includegraphics[width=0.9\textwidth]{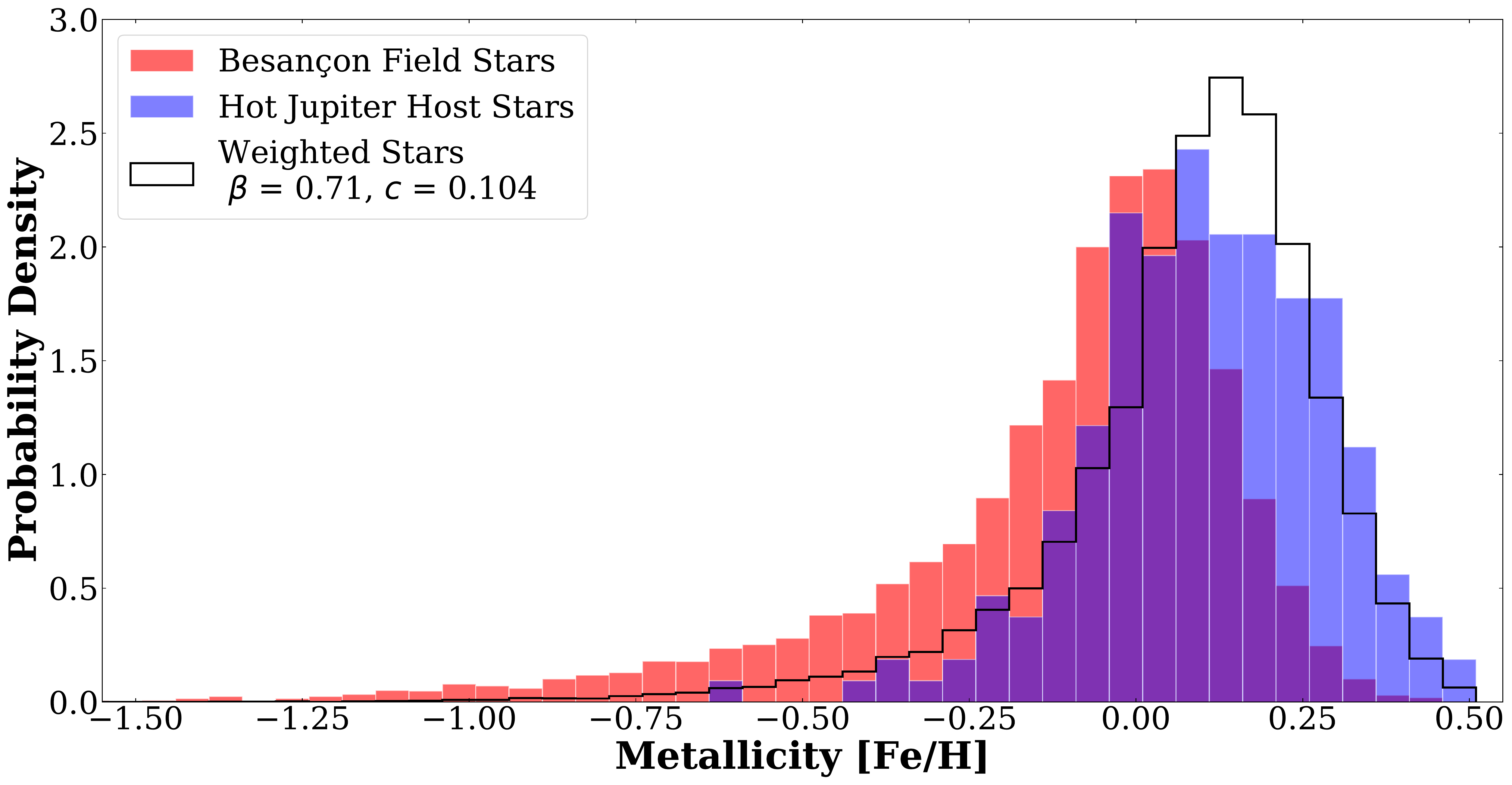}
    \caption{Metallicity distributions for field stars simulated with the \bes\ model (red), and \hJ\ host stars from our sample with metallicities taken from \sweet\ (blue). A weighted sample distribution is also displayed (black outline), which corresponds to weighting the simulated field stars by applying Equation \ref{eq:myequ} (using the specific values of $\beta$ and $c$ estimated by the MCMC), as if they all hosted hot Jupiters.}
    \label{fig:histogram}
\end{figure*}

\subsection{The planet-metallicity correlation} \label{subsec:fitting}

In the seminal work on the planet-metallicity correlation of \citet{Valenti2005}, it was shown that the probability of finding a giant planet rises sharply as a function of the metallicity of the star. \citet{Valenti2005} utilises a power law of the functional form  
\begin{equation}
    f(\rm{[Fe/H]}) \propto 10^{\beta \rm{[Fe/H]}}
\label{eq:fischer}
\end{equation}
to relate $f$, the fraction of stars with giant planets, to metallicity (\feh), where $\beta$ is the index of the power law.  We adopt this same formalism in quantifying the planet-metallicity correlation for hot Jupiters from our sample.  We note that unlike \citet{Valenti2005} and many subsequent surveys, we are not sensitive to probing the absolute occurrence rates of giant planets.  Since we are simulating the stars from which our hot Jupiter hosts stars are drawn, we cannot determine the absolute occurrence rate of hot Jupiters. Instead, we solely probe the dependence of hot Jupiter occurrence on metallicity.

As discussed in Section~\ref{sec:sample}, is it widely known that different analysis methods used to derive metallicity from optical spectra can introduce significant offsets in measured metallicity \citep{Torres2012, Petigura2018}.  This issue is equally true when comparing the metallicities of our hot Jupiter hosts from \sweet\ \citep{Santos2013} to the metallicities of our simulated population of stars from the \bes\ Galaxy Model \citep{Robin2003}.  In order to address this issue, we need to allow for an offset between the \sweet\ metallicities and the \bes\ metallicities.  We do this by adding a offset term, $c$, to Equation~\ref{eq:fischer} as follows:
\begin{equation}
    f(M,\rm{[Fe/H]}) \propto 10^{\beta (\rm{[Fe/H]} + c)}.
    \label{eq:myequ}
\end{equation}

Since we have a large sample of hot Jupiter hosts, and the offset is a linear shift between the \sweet\ and \bes\ metallicities, we can simply fit for $c$ and $\beta$ simultaneously when comparing the populations of hot Jupiter hosts to the simulated field stars.

We again use a KS test to compare the metallicities of the \hJ\ host star population to the \bes\ simulated population, and explore the parameter space of $\{\beta,c\}$ in Equation~\ref{eq:myequ} using the MCMC sampler \textsc{emcee} \citep{DFM2013}.  The log of the p-value output of the KS test was taken as the log likelihood at each step in the chain - we are maximising the p-value as a higher p-value indicates that the values for $\beta$ and $c$ are a better fit.

After a preliminary search over the $\{\beta,c\}$ parameter space, uniform priors were placed on both $\beta$ and $c$, but they were restricted to the ranges $0 \leq \beta \leq 1.8$ and $0 \leq c \leq 0.2$. This process was run for 5000 steps after an initial 500 that were discarded as burn-in.  The results are set out in Section~\ref{sec:results} below.


\section{Results}
\label{sec:results}
We find that for our uniform sample of 217 \hJs, there is a clear difference in metallicity between our \hJ\ host stars and the simulated field star comparison sample. Figure~\ref{fig:histogram} displays the metallicity distributions for our simulated field stars (in red) compared to our \hJ\ host stars (in blue).  The histograms are clearly distinct in terms of their distributions, with the simulated field stars being less metal-rich than the \hJ\ hosts. Specifically, the mean metallicity of the simulated field stars is \feh=-0.115 $\pm$ 0.003~dex, while the \hJ\ host stars is \feh=0.100 $\pm$ 0.012~dex, where the error is given as the standard error of the mean. This gives a significant metallicity difference of 0.215 dex.  A KS test comparing the distributions gives a statistic of 0.35 and a p-value of $1.47 \times 10^{-23}$, allowing us to reject the null hypothesis: these 2 samples are not drawn from the same population. 

\begin{figure}
	\includegraphics[width=\columnwidth]{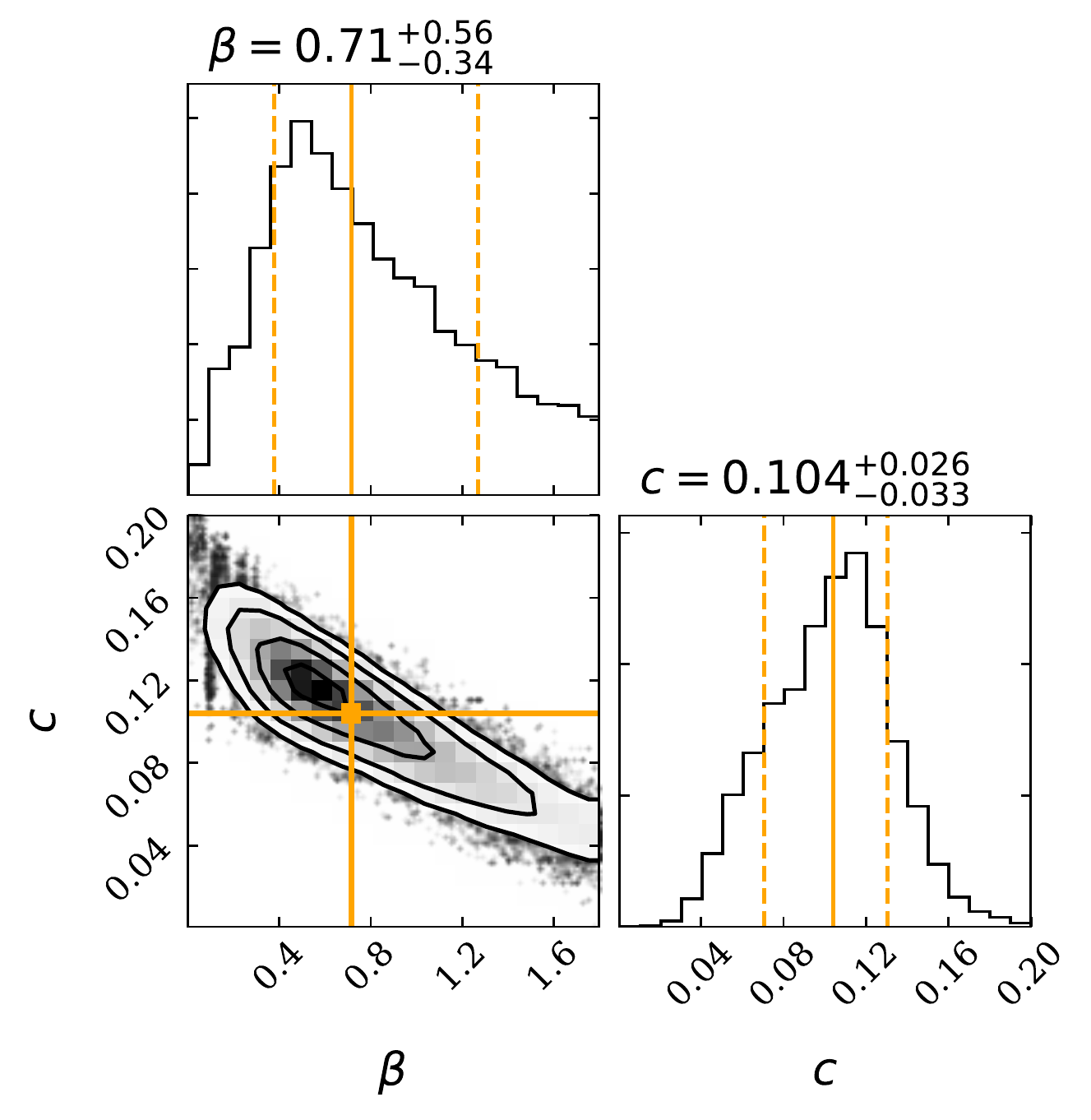}
    \caption{Corner plot displaying a 2D contour plot and 1D histograms of the samples drawn during the MCMC exploration of the parameter space. The solid orange lines show the median values, and the dashed orange lines show the lower and upper uncertainties using the 16th and 84th percentiles. The histogram titles display the median and $\pm 1 \sigma$ uncertainties for each parameter.
    This plot was made using the \textsc{corner.py} code \citep{DFM2016}. }
    \label{fig:corner_a0}
\end{figure}

The exploration of the $\{\beta,c\}$ parameter space allows us to disentangle the degree to which this metallicity difference is due to a systematic metallicity offset between \sweet\ and \bes, or an intrinsic planet-metallicity correlation for \hJs.  Figure~\ref{fig:corner_a0} shows the corner plot of the samples drawn in the MCMC exploration of the $\{\beta,c\}$ parameter space described in Section~\ref{sec:analysis}. From these samples, we estimate values for $\beta$ and $c$ of $0.71^{+0.56}_{-0.34}$ and $0.104^{+0.026}_{-0.033}$ respectively. These were estimated using the 16th, 50th, and 84th percentiles.

In Fig.~\ref{fig:histogram} we show the expectation for the metallicity distribution of the simulated field stars weighted as if they all hosted \hJs\ (black outline) - i.e. applying Equation~\ref{eq:myequ} with our best fit $\beta$ and $c$ from the MCMC exploration ($\beta=0.71, c=0.104$). We see the weighted sample distribution closely approximates the \hJ\ host star distribution.
The KS test result for the weighted sample distribution with the best fit values of $\beta$ and $c$ gives a statistic of 0.062 and a p-value of 0.383. This shows that this distribution and the \hJ\ host star distribution are indistinguishable.


\section{Discussion}
\label{sec:discussion}
Our key result is that \hJs\ show a planet-metallicity correlation that follows a power law with $\beta=0.71^{+0.56}_{-0.34}$.  In this Section we compare this to previous studies, examine any potential biases in our statistic sample, and discuss the implications of our results in terms of the formation and migration of hot Jupiters.  

\subsection{Comparisons with previous studies}
\citet{Valenti2005} studied a sample of 1040 FGK stars from a long term, homogeneous radial velocity survey, and found that $\beta=2$ in the regime of giant planets with orbital periods $<4$\,years.  No uncertainties are placed on their result.

\citet{Johnson2010} also studied the giant planet-metallicity correlation, this time for a sample of 1194 stars covering a wider stellar mass range of AFGKM stars drawn from a combination of the Keck M Dwarf Survey, the original SPOCS catalogue from \citet{Valenti2005}, and the SPOCS IV catalogue. 
They found a value of $\beta=1.2\pm0.2$, which is slightly higher but fully consistent with our result.  

\citet{Schlaufman2014} used a sample of 620 FGK stars, 44 of which host at least one giant planet, from the HARPS GTO program (taken from \citet{Adibekyan2012}), using logistic regression to derive a $\beta$ value of $2.3 \pm 0.4$.  Interestingly, this result is in good agreement with the \citet{Fischer2005} result, but not with the result of \citet{Johnson2010} or with our result. 

While all of the above results are from radial velocity surveys, there have also been previous attempts to calculate $\beta$ from transit surveys, in particular from the \kepler\ survey \citep{Borucki2010}. \citet{Guo2017}  and \citet{Petigura2018} both evaluate $\beta$ for the population of 14 hot Jupiters in the \kepler\ data. \citet{Guo2017} find a value of $\beta=2.1 \pm 0.7$, consistent with radial velocity results, while \citet{Petigura2018} find a value of $\beta=3.4^{+0.9}_{-0.8}$, which is higher than previous studies.  The low number of hot Jupiters from \kepler, coupled with the complex targeted nature of the survey (c.f. the untargeted surveys used in our sample), means that these results need to be approached with some caution.  

Our result ($\beta=0.71^{+0.56}_{-0.34}$) confirms the giant planet-metallicity correlation seen in previous studies, but suggests that it is marginally \textit{weaker} for hot Jupiters than it is for the longer period giant planets such as in the survey outlined above.  We summarise our result and the previous results in Table~\ref{tab:betas}.

Our $\beta$ value is lower than all previously published results in Table~\ref{tab:betas}. We are within $1 \sigma$ of the result of \citet{Johnson2010}; however, if we accounted for the mass dependency in our calculation as they have, we would expect our value for $\beta$ to decrease further (though not significantly so, as the stellar masses in our sample have a range of only $\approx 1 \msun$). We are 2.32 and 2.31$\sigma$ from \citet{Valenti2005} and \citet{Schlaufman2014}, the two other results from RV surveys, respectively. While these hint at a difference in the strength of the correlation between cool and hot Jupiters, we are also 1.55$\sigma$ removed from \citet{Guo2017}, and 2.75$\sigma$ from \citet{Petigura2018}, the two hot Jupiter specific studies. Though again it should be noted that both hot Jupiter studies have a very small sample size. 

\begin{table}
	\centering
	\caption{$\beta$ values}
	\label{tab:betas}
	\begin{tabular}{llc} 
Reference & Stellar Sample & $\beta$ \\
\hline
\citet{Valenti2005} & 1040 FGK stars & 2.0 \\
 & (RV survey) & \\
\citet{Johnson2010} &  1194 AFGKM stars  & $1.2 \pm 0.2$  \\
&   (RV survey) &  \\
\citet{Schlaufman2014} & 620 FGK stars & $2.3 \pm 0.4$ \\
&  (RV survey) &  \\
\citet{Guo2017} & 13 Kepler hot\ & $2.1 \pm 0.7$ \\
 &  Jupiter hosts &  \\
\citet{Petigura2018} & 14 Kepler hot\  & $3.4^{+0.9}_{-0.8}$ \\
 &  Jupiter hosts &  \\
This work & 217 \hJ\  & $0.71^{+0.56}_{-0.34}$ \\
 & hosts &  \\
		\hline
	\end{tabular}
\end{table}

\subsection{Potential biases}
\label{sec:bias}

The metallicity offset ($c$) is needed to calibrate between the metallicities in \bes\ and \sweet, but adds an extra degree of uncertainty compared with a survey that has a uniformly determined set of metallicities for both hot Jupiter hosts and field stars.  However, the metallicity offset appears fairly well constrained from the sample distribution in Fig. \ref{fig:corner_a0}, and is relatively small in comparison to the overall spread of metallicities (c.f. Fig. \ref{fig:histogram}).  The metallicity offset does correlate with $\beta$ (see Fig. \ref{fig:corner_a0}), which results in a relatively large and slightly asymmetric uncertainty on $\beta$.

There is also a correlation between the radius of a star and its metallicity: the increase in opacity with the presence of metals results in the star having a larger radius. As transit depth decreases with the square of a star's radius, planets would be more difficult to detect around higher metallicity stars via the transit method. This would act to decrease our value of $\beta$, but it has been found by \citet{Petigura2018} that planet detectability does not significantly depend on stellar metallicity.

RV surveys will preferentially find planets around metal rich stars as it is easier to perform the method when there are stronger and/or more metal lines present in the host star spectra. We do not expect the detection of a planet via the transit method to depend significantly on metallicity of the host star - and this is one of the advantages of using a sample of hot Jupiter planets from transit survey discoveries. However, it should be noted that confirmation of planets from transit surveys is based on RV follow-up, which will still be subject to the bias described above.

\subsection{Hot Jupiter Formation and Migration}
\label{sec:formation}

Hot Jupiters were an unexpected discovery, given how close-in they are to their host stars and that they have no solar system analogue. Due to the lack of disk mass close to a star, in-situ formation was thought to be unlikely; instead, it has been posited that hot Jupiters form far out from their star, beyond the snowline, and then undergo inward migration after or during their formation. Core accretion, supported by the planet-metallicity correlation, together with disk-driven migration and interactions with planetary companions when the hot Jupiter is misaligned with the stellar rotation axis \citep[e.g.][]{Dawson2013} are currently thought to be the main mechanisms producing hot Jupiters. In-situ formation has, however, been recently reconsidered to be a possibility \citep[e.g][]{Boley2016}. 

\citet{Maldonado2018} makes the assumption that hot and cool Jupiters would have similar chemical properties if hot Jupiters were formed at large distances from their star and then migrate inwards, but they find that hot and cool Jupiters have different properties, and that they are two distinct populations. Perhaps they have different formation methods, or perhaps hot Jupiter migration is a metallicity dependent process. \citet{Maldonado2018} argues that the latter is unlikely, as it would not be expected that migration would change the abundance of the host star. 

A number of studies examine the relationship between metallicity of a host star and the orbital period of different planet types in the system, including giant planets. The result of \citet{Narang2018} finds no difference in metallicity with orbital period for giant planets. \citet{Adibekyan2013} find that, from $\sim$ 10 \mearth\ to $\sim$ 4 \mjup, planets in metal-poor systems have longer periods than those in metal-rich systems. They suggest this may be due to planets in a metal-poor disk forming further out and/or undergoing later and thus less migration as they take longer to form. \citet{Mulders2016} finds that, while occurrence rate of hot rocky exoplanets within a 10 day orbital period increases with metallicity, hot gas giants exhibit no significant relationship between metallicity and orbital period. 

Our result that hot Jupiters preferentially orbit metal-rich stars is in agreement with all past results on the planet-metallicity correlation, and is more evidence towards the core accretion model of formation. As our value for $\beta$ is consistent with past RV survey results (though marginally weaker), it suggests that hot and cool Jupiters may form in the same way, and that their migration is different. The nature of this correlation might be an indication against in-situ formation - you would expect in-situ formation to be enabled by higher amounts of metals compared to systems which form planets further out, but our result does not indicate an comparative increase in the metallicity of hot Jupiter systems over longer period gas giants.


\section{Conclusion}
\label{sec:conclusion}

We have examined the giant planet-metallicity correlation using the host stars of hot Jupiter planets, based on a sample of 217 hot Jupiters taken from the transit surveys WASP, HATNet, HATSouth, KELT, XO and TrES, with metallicities taken from \sweet. We compare these to a population of field stars simulated with the \bes\ Galaxy model, and find a clear difference in their metallicity distributions, with the hot Jupiter hosts being more metal rich.  We use the formalism of \citet{Valenti2005} (Equation \ref{eq:fischer}) and find $\beta=0.71^{+0.56}_{-0.34}$.  This result is lower, but consistent to within uncertainties, to $\beta$ values derived from radial velocity surveys that probe much longer period giant planets \citep[e.g.][]{Valenti2005,Johnson2010}.  We conclude that this is strong evidence to suggest that the population of hot Jupiter giant planets is not a distinct population, but is drawn from the same population as giant planets on longer orbital periods.  This result will be able to be confirmed by the complete set of hot Jupiter planets orbiting bright stars that should arise from the TESS mission \citep{Ricker2015}, in conjugation with a more complete and consistent survey of stellar metallicities.

\section*{Acknowledgements}

We thank the anonymous referee for their constructive comments that have contributed to the quality of this paper. We thank George King for his help in setting up the MCMC, and we thank both George King and David Brown for their helpful comments that improved the initial manuscript. This research has made use of the NASA Exoplanet Archive, which is operated by the California Institute of Technology, under contract with the National Aeronautics and Space Administration under the Exoplanet Exploration Program.


\bibliographystyle{mnras}
\bibliography{ms} 






\bsp	
\label{lastpage}
\end{document}